\def\year{2024}\relax
\documentclass[letterpaper]{article} 
\usepackage{aaai22}  
\usepackage{times}  
\usepackage{helvet}  
\usepackage{courier}  
\usepackage[hyphens]{url}  
\usepackage{graphicx} 
\usepackage{natbib}  
\usepackage{caption} 
\usepackage{subcaption}
\DeclareCaptionStyle{ruled}{labelfont=normalfont,labelsep=colon,strut=off} 
\usepackage{algorithm}
\usepackage{algorithmic}

\usepackage{xcolor}

\usepackage{xcolor}
\usepackage{hyperref}
\usepackage{color, colortbl}
\definecolor{Gray}{gray}{0.95}
\usepackage{wrapfig}
\usepackage[font=normal,labelfont=bf]{caption}

\usepackage{booktabs} 
\usepackage{caption} 
\usepackage{newfloat}
\usepackage{listings}
\floatstyle{ruled}
\newfloat{listing}{tb}{lst}{}
\floatname{listing}{Listing}

\title{Election Polls on Social Media: Prevalence, Biases, and Voter Fraud Beliefs}
\author{
    Stephen Scarano\textsuperscript{\rm 1 \dag}, Vijayalakshmi Vasudevan\textsuperscript{\rm 1}, Mattia Samory\textsuperscript{\rm 2 \S}, Kai-Cheng Yang\textsuperscript{\rm 3},\\ JungHwan Yang\textsuperscript{\rm 4 \P}, and Przemyslaw A. Grabowicz\textsuperscript{\rm 1 \ddag}
}
\affiliations{
    \textsuperscript{\rm 1} University of Massachusetts Amherst, USA\\
    \textsuperscript{\rm 2} Sapienza University of Rome, IT\\
    \textsuperscript{\rm 3} Network Science Institute, Northeastern University, USA\\
    \textsuperscript{\rm 4} University of Illinois Urbana-Champaign, USA \\
    \textsuperscript{\rm \dag} sscarano@umass.edu,
    \textsuperscript{\rm \ddag} grabowicz@cs.umass.edu,
    \textsuperscript{\rm \S} mattia.samory@uniroma1.it,
    \textsuperscript{\rm \P} junghwan@illinois.edu 
}

\begin{document}

\maketitle

\begin{abstract}
Social media platforms allow users to create polls to gather public opinion on diverse topics.
However, we know little about what such polls are used for and how reliable they are, especially in significant contexts like elections.
Focusing on the 2020 presidential elections in the U.S., this study shows that outcomes of election polls on Twitter deviate from election results despite their prevalence.
Leveraging demographic inference and statistical analysis, we find that Twitter polls are disproportionately authored by older males and exhibit a large bias towards candidate Donald Trump relative to representative mainstream polls.
We investigate potential sources of biased outcomes from the point of view of inauthentic, automated, and counter-normative behavior. Using social media experiments and interviews with poll authors, we identify inconsistencies between public vote counts and those privately visible to poll authors, with the gap potentially attributable to purchased votes.
We also find that Twitter accounts participating in election polls are more likely to be bots, and election poll outcomes tend to be more biased, \textit{before} the election day than \textit{after}.
Finally, we identify instances of polls spreading voter fraud conspiracy theories and estimate that a couple thousand of such polls were posted in 2020.
The study discusses the implications of biased election polls in the context of transparency and accountability of social media platforms.
 
\end{abstract}

\maketitle

\section{Introduction}\label{sec:intro}

The advent of social media and the Internet has revolutionized the landscape of political discourse.
The unique technological affordances provided by popular social media platforms, such as \textit{Twitter}\footnote{Throughout the paper, we will use \textit{Twitter} to refer to \textit{X}.} (now known as \textit{X}) and \textit{Facebook}, enable people to become commentators and active distributors of information rather than mere consumers of it.
Consequently, social media have garnered significant attention from the public, media, and political elites, who now use these online discussions to gauge the public's views on critical issues and utilize those insights for political campaigning~\cite{harfoush2009yes, hughes2010obama}.

However, social media may not accurately represent public opinion. Instead, it disproportionately reflects the views expressed by a reactive, polarized audience \cite{Zhang2022}. Additionally, the demographic profile of Twitter users is non-representative \cite{wojcik2019twitterpop}, and the presence of bots, astroturfing accounts \cite{keller2020campaign, 10.1145/2818717}, and foreign influence~\cite{Schoch2022} further complicates the situation. 
The overrepresentation of certain perspectives, driven by a minority of users, can skew social media users' perceptions of reality and bias political processes.

Public opinion research has shown that opinion polls can influence how people perceive public opinion and shape how they form their own opinions ~\cite{Lang1984, Marsh1985, Morwitz1996}. Politicians and news outlets may also place greater emphasis on the issues that the public cares more about, further shaping policy decisions~\cite{Arnesen2017}. Given the significance of polling in policy-making, it is important to note that
Twitter, a leading social media platform, introduced polls in 2015,\footnote{\url{https://blog.twitter.com/official/en_us/a/2015/introducing-twitter-polls.html}} a year before the 2016 U.S. presidential elections.

\begin{figure}
    \centering
    \includegraphics[width=0.95\linewidth]{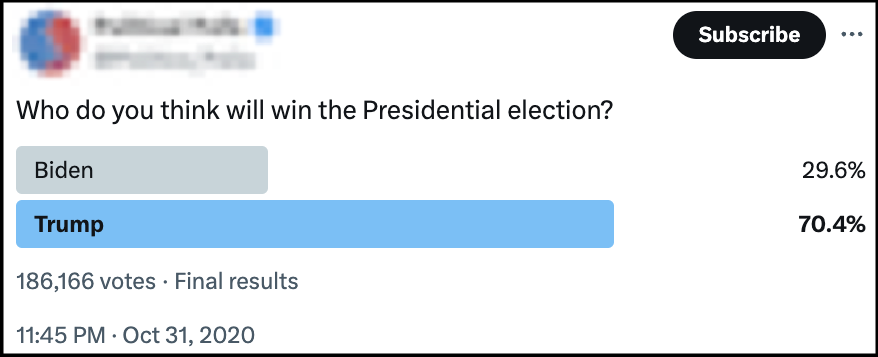}
    \caption{An example of a social poll from our dataset.}
    \label{fig:poll-example}
\end{figure}

Social polls are straw polls that any social media user can create and share with other users on the platform (see example in Figure~\ref{fig:poll-example}). As a social media feature, they have become remarkably successful. In 2022, Elon Musk 
notably used this feature to make key business decisions, such as changing Twitter's CEO \citep{Mehta2022}. Twitter polls have gained popularity due to their wide reach, quick turnaround time, ease of use, and low cost. It is common to see popular Twitter polls that amass millions of votes.

However, no research has yet examined the extent to which social polls are used in political campaigns and their role in these campaigns. To address this research gap, the first goal of this study is to explore the following research question.

\vspace{2pt}
\textit{\textbf{RQ1}: How many Twitter polls related to the U.S. presidential elections were published in 2016 and 2020? Did the number of such polls grow between the two election cycles?}
\vspace{2pt}

Twitter polls inherently lack scientific rigor due to the absence of systematic sampling and demographic information about the respondents, resulting in potential bias in poll outcomes. 
Despite this, little attention has been paid to understanding the biases in social polls and their broader impact. Therefore, we study the biases in social polls, addressing the following set of related research questions.

\vspace{2pt}
\textit{\textbf{RQ2.1}: How do Twitter election poll outcomes deviate from election results and traditional poll outcomes?}
\vspace{2pt}

After establishing that Twitter poll outcomes are biased, we attempt to reason about the potential sources of these biases. First, we note that Twitter poll votes can be purchased from online vendors.

\vspace{2pt}
\textit{\textbf{RQ2.2}: How does Twitter account for purchased votes? Did such questionable votes appear in polls related to the 2020 U.S. presidential election?}
\vspace{2pt}

Second, election polls may have suspicious participants, such as bots, foreign accounts, or hyperactive users. We compare the percentage of Twitter users that are identified as automated them with a reference set, before and after elections.

\vspace{2pt}
\textit{\textbf{RQ2.3}: Did the voting activity of bots, foreign accounts, and hyperactive users change after the election day?}
\vspace{2pt}

Third, poll outcomes may also be influenced by legitimate demographic and political differences among users who participate in the polls.

\vspace{2pt}
\textit{\textbf{RQ2.4}: What are the characteristics of users engaging with Twitter election polls in comparison to U.S. voters and random non-political polls? }
\vspace{2pt}

Finally, we compare these three factors and study their relationship with biases in  poll outcomes.

\vspace{2pt}
\textit{\textbf{RQ2.5}: Which attributes of users correlate with biases in Twitter election poll outcomes? }
\vspace{2pt}

Individuals wary of institutions may perceive social polls as more legitimate and correct measures of public opinion than mainstream polls, in that the former are borne by the initiative of any social media user. This perception is especially problematic, considering that of the major issues during the 2020 U.S. presidential elections was the allegation of voter fraud by presidential candidate Donald Trump and his supporters~\cite{pennycook2021false}. The issue has remained so prevalent that, as of the writing of this manuscript, it is still influencing the discourse surrounding the 2024 U.S. presidential election. 
At the core of the voter fraud conspiracy theory is the belief that the official election results do not represent public opinion. 
We focus on Twitter polls as a potential source of such beliefs. 

In particular, while widespread biases in the outcomes of Twitter election polls may misrepresent the popular support for presidential candidates, it is not clear whether such biased polls are indeed \textit{explicitly} and \textit{intentionally} used to support voter fraud beliefs surrounding the 2020 U.S. presidential election. 

\vspace{2pt}
\textit{\textbf{RQ3}: How many Twitter election polls explicitly question election integrity and the accuracy of mainstream polls? How many users interact with such conspiratorial polls?}
\vspace{2pt}

We conclude the study by discussing the broader implications of biased election polls, including transparency and accountability of social media platforms such as Twitter.

\section{Related Work}

We analyze thousands of Twitter polls related to U.S. presidential elections. To our knowledge, such polls have not been studied to date. 
To provide a proper foundation for our study, we begin with a review of existing literature. 

\subsubsection{Potential Biases in Social Polls} \label{related-work-socpol}

Public opinion, defined as an aggregate of individual opinions \cite{Price1992}, is a crucial part of a well-functioning democracy. During the election, understanding public opinion becomes particularly important as it offers insight into the voters' support towards political candidates, which can inform campaign strategies. One of the most popular ways to assess public opinion is through survey-based opinion polls \cite{Price1992}. To accurately measure public opinion using polls, it is crucial to draw an unbiased and representative sample of the target population (i.e., electorates) \cite{Squire1988}. 

Over the past decade, social media have emerged as significant platforms for individuals to express their opinions on various social issues. However, although Twitter users constantly produce millions of tweets every hour, these tweets may not accurately represent public opinion. 
This discrepancy arises due to several factors. First, the opinions of specific demographics can be over-represented because (i) Twitter users are more likely to be male and young \citep{mislove2011understanding, wojcik2019twitterpop}, (ii) nearly 80\% of tweets on the platform are published by a mere 10\% of the most active users, and (iii) politically-interested users are non-representative of all users \citep{hughes2019national,hughes2021small}.
Additionally, the prevalence of bot accounts, astroturfing campaigns, and artificial likes and comments, can distort various metrics from Twitter, including social polls \citep{keller2020campaign, Ferrara2016}. 

In the following section, we discuss related work on two potential sources of bias: inauthentic behavior, which may make active attempts to manipulate the visibility of certain views, and the lack of representativeness of the users who engage in social polling compared to the actual population of voters, which may over-emphasize specific political views.

\subsubsection{Digital Manipulation} \label{related-work-astroturf}

Inauthentic social media activities, such as bots and astroturfing, can contribute to social media biases. This phenomenon was famously manifested in Russia's \textit{Internet Research Agency} (IRA) intervention in the 2016 U.S. Presidential election, where many organized or sponsored inauthentic social media operations are uncovered and documented \citep{keller2020campaign, Schoch2022}. Similarly, research has found that numerous bot accounts generate a significant amount of social media posts \citep{Ferrara2016}. Furthermore, the prevalence of bot accounts on Twitter has increased since Elon Musk's acquisition of the company in 2022 \citep{hickey2023auditing}. 

Tools for manipulating social media discourse, now broadly accessible via online services, also have expanded their reach beyond governmental institutions to everyday consumers \citep{alrawi2020ira}. These external influences have the potential to distort the representation of public opinion, particularly as politically motivated campaigns deliberately exploit them to sway the course of social media dialogues \citep{Schoch2022}.

Existing studies suggest that inauthentic social media accounts can influence Twitter discourse beyond merely generating messages or retweeting. For instance, in the 2012 South Korean election, where national intelligence agents conducted an extensive disinformation campaign supporting one of the presidential candidates by manually operating multiple Twitter accounts. The agents tried to make these astroturfing accounts appear genuine while amplifying messages through retweets, likes, and cross-posting \citep{keller2020campaign}. Similarly, the \texttt{\#YaMeCanse} online protest movement in Mexico demonstrated coordinated influence, where astroturfing accounts manipulated the hashtag to stifle discourse \citep{su2016protests}. In the \texttt{\#YaMeCanse} movement case, researchers using \textit{Botometer}, a publicly available machine learning algorithm, found that only 10-14\% of the accounts were suspected bots, while the rest appeared to be operated potentially by humans.

While prior research has primarily focused on analyzing social media messages and sharing patterns \citep{keller2020campaign, Schoch2022, su2016protests, Zhang2022, Ferrara2016}, it has largely overlooked potential social polls and their potential manipulation. This study uniquely contributes to the field by examining inauthentic activities and biases in social polls. 

\subsubsection{Demographics of Social Media Users}

One of the main sources of biases in social polls is the unrepresentative demographic traits of social media users. Social media users do not accurately reflect the demographic makeup and location distribution of the general population. Audiences on Twitter and Facebook are significantly younger and lean more toward the political left \citep{mellon2017population, wojcik2019twitterpop}. Facebook users skew towards a female demographic, while Twitter users are biased towards men \citep{mislove2011understanding, mellon2017population, wojcik2019twitterpop}. In addition, social media users generally tend to over-represent the wealthier and more educated population \citep{wojcik2019twitterpop}. 

Another important factor contributing to potential biases in social polls is the political interest and ideology of social media users. 
According to data from the Pew Research Center, social media users in the United States tend to skew more liberal in their political affiliations compared to the general population. For instance, 36\% identify as Democrats among social media users, whereas the corresponding figure in the U.S. general population is 30\% \citep{wojcik2019twitterpop}. 

\subsubsection{Conspiracy Beliefs via Polls}
The creation and posting of opinion polls on social media are often driven by the prevalence of conspiratorial beliefs in contemporary politics. Voter fraud is one of the prominent conspiracy theories. Research has found voter fraud beliefs regarding mail-in voter fraud and the manipulation of social media were popular during U.S. elections \cite{Benkler2020MailInVF, Ferrara_Chang_Chen_Muric_Patel_2020}. These studies highlight how such misinformation has the potential to significantly undermine the integrity and proper functioning of democratic systems.

A potential source of voter fraud belief is survey questions. Survey questions that incorporate conspiracy theories can inadvertently contribute to mistrust in mainstream information \cite{Clifford2023}. This effect is largely attributed to the ``question wording'' and ``panel conditioning,'' a phenomenon where exposure to specific ideas within surveys shapes respondents' beliefs \cite{Kalton1982, Das2011}. This is particularly evident when similar questions are repeated in multiple surveys, where respondents tend to learn from the survey questions they have previously answered.

Polls that elude conspiracy beliefs, such as voter fraud claims, might be particularly influential because they often provide simplified, alternative explanations for complex events. This impact is more pronounced when respondents have limited initial information on the topic \cite{Marchlewska2018}, hold strong pre-existing political beliefs \cite{Pennycook2018}, or have a predisposition towards conspiratorial thinking \cite{Uscinski2016}. Thus, given the prominence of conspiratorial narratives around election integrity, the design and content of survey questions, especially those involving conspiracy theories about the electoral process or legitimacy of information, can promote mistrust in the government and the overall democratic process in society \cite{pennycook2021false}. However, the extent to which such polls exist on Twitter is largely unknown.

\begin{table}
\centering
\resizebox{\columnwidth}{!}{
\begin{tabular}{| l | c | c | c |}
\hline
\rowcolor{Gray}
\textbf{Year} & \textbf{2016} & \multicolumn{2}{c|}{\textbf{2020}}  \\
\hline
\rowcolor{Gray}
\textbf{Dataset source} &  \textbf{Query} & \textbf{Query}& \textbf{Decahose} \\
\hline
\rowcolor{Gray}
\textbf{Utilized to address} & \textbf{RQ1} & \textbf{RQ1, 2} & \textbf{RQ1, 2.3, 3}\\
\hline
Polls related to elections & 1,759  & 4,900 & 12,990 \\
Polls gauging support & 510 & 1,440  & \\
\hline
Engagement counts & + & + & + \\
User demographics & Not shown & + & + \\
User political affiliation & Not shown & + &  \\
Poll author survey &  & + &  \\
\hline
\end{tabular}
}\caption{Summary of our three datasets of Twitter polls showing which dataset we use to answer each of our Research Questions (RQs).
The table shows which information we miss for the 2020 Decahose and the 2016 datasets. We are unable to acquire the missing data due to Twitter's removal of academic API access. 
}
\label{tab:datasets}
\end{table}

\section{Datasets and Twitter Restrictions}
\label{sec:datasets}

We collected three large datasets of Twitter polls related to the 2020 and 2016 U.S. presidential elections (see Table~\ref{tab:datasets}). 
All datasets contain polls related to elections in the sense that they have the respective two main presidential candidates among poll options. 
To estimate the growth and overall prevalence of election polls (RQ1), we use the dataset for 2016 and the two datasets for 2020. 
Throughout the remainder of the manuscript, we use only the 2020 datasets, since 
Twitter API restrictions prevented us from collecting a more complete 2016 dataset.
Below, we describe the datasets and the implications of Twitter API restrictions.

\subsection{Datasets}

\subsubsection{Election Query Polls}
Provided access to the Twitter API v2, we collected election polls from the periods of 1/1/2020--11/30/2020 and 1/1/2016--11/30/2016 by making full-archive searches for the queries ``\texttt{vote AND (trump OR biden)}'' and ``\texttt{vote AND (trump OR clinton)}'' respectively. 
This step resulted in a vast number of polls. 
To identify polls related to presidential elections, we focused on the polls that explicitly mention respective presidential candidates among poll options (``\texttt{trump OR donald}'' and ``\texttt{joe OR biden}'' or ``\texttt{hillary OR clinton}''). 
This resulted in 4,900 polls. We refer to these polls as query polls to distinguish them from random polls described next.

\subsubsection{Random Election Polls}
We have access to the Decahose stream, a 10\% random sample of all tweets produced on Twitter.
We drop from this dataset all retweets to avoid bias towards polls with multiple retweets.\footnote{Given that each retweet counts as a tweet, polls that have multiple retweets are more likely to appear in Decahose than by chance. If we did not disregard them, our sample of polls would be biased towards popular polls, would be non-representative of all polls, and our estimates, e.g., of the total numbers of polls, would be incorrect.} 
Then, to identify polls related to elections we apply the same filtering step as for the query polls, described above, i.e., we select polls that mention ``\texttt{trump}'' and ``\texttt{biden}'' among poll options.
The main advantage of this dataset is that it is a random 10\% sample (without retweets) of all polls, whereas query dataset includes only the polls that mention the word ``vote'' and either ``trump'' or ``biden'' in its main text (not poll options), and we do not know what fraction of all polls they represent.
However, the Decahose sample of polls has a few crucial limitations. First, the Decahose stream does not contain any polls that were published in 2016, since the polling feature was introduced to Twitter just one year earlier. 
Second, due to Twitter API restrictions, we are unable to complete it, as we explain in the next subsection. 

\subsubsection{Mainstream and Exit Polls}
To make comparisons with the Twitter polls, we use 192 mainstream polls from 2020 aggregated by \textit{FiveThirtyEight},\footnote{\url{https://projects.fivethirtyeight.com/polls}} and national exit poll data distributed by the \textit{Roper Center} \citep{NEP2020}.

\subsubsection{Polls Gauging Support for Presidential Candidates}
To compare the results of Twitter election polls with election results and mainstream polls (RQ2.1) and to explain biases in Twitter poll outcomes (RQ2.5), we must identify Twitter polls that gauge popular support for the U.S. presidential candidates.
For instance, polls that ask ``Who won the last debate?'' are related to the elections and appear in our datasets but do not gauge the overall (rather than momentary) support for the presidential candidates.
To identify relevant polls, we manually inspected all of the query polls from 2016 and 2020. To this end, we developed a labeling guideline that defines polls gauging support for presidential candidates as either (i) directly asking for voting preferences (e.g., ``Who has your vote?''), or (ii) asking for election predictions (e.g., ``Who do you think will win the presidential election vote?'').
This approach identified a total of 1,440 Twitter polls gauging support for the 2020 presidential candidates.
To estimate the inter-rated agreement, a subset of $194$ polls was labeled by two trained coders. They have achieved an almost perfect inter-rater agreement on this set of items, measured with Cohen's kappa, of $0.914$ $(p<0.001)$. 

\subsubsection{Classification and Validation of User Traits}
For each poll from our three datasets, we collect all retweeters and favoriters.
For the polls from the query-based datasets, we collect followees of all poll authors, necessary for political affiliation identification.
For each poll author, retweeter, and favoriter, we use state-of-the-art classifiers to identify that user's bot score, age, gender, organization status, and political affiliation.\footnote{We did not label political affiliation for 2016 query and 2020 Deacahose polls, because we were not able to gather followee information of users due to Twitter API removal of Academic API access.} We describe the classifiers in respective methods sections focused on these user traits.

To validate the classifier labels we asked one human coder to review the estimated outcome of those four attributes from a random set of 239 Twitter accounts participating in the vote query polls. 
To simplify the process, we showed the machine-determined attributes to the coder and asked them to determine whether they agreed with those classifications or not. 
Based on the coder's validation, our methods achieved approximately 93\% accuracy in distinguishing between organizational and personal accounts, 91\% accuracy in assessing bot-likeness, 93\% accuracy in estimating political ideology, 91\% accuracy in estimating the age of the account holders, and 88\% accuracy in classifying the gender of the account holders.

\subsection{Use and Limitations of Datasets}

\subsubsection{Impact of Twitter API Restrictions}
Due to the closure of Academic API by Twitter in 2023 and changes to the Twitter user interface at the end of 2022, we were not able to (Table~\ref{tab:datasets}): (a) conduct a survey of poll authors neither for the 2016 query polls nor for the Decahose polls, and (b) collect the information about user followees necessary to infer political affiliations of users interacting with the Decahose polls. We provide details of these two issues in the sections addressing RQ2.2 and RQ2.4, respectively.

\subsubsection{Datasets vs. Research Questions}
In consequence, as our main dataset, we use the 2020 query dataset, which is our most complete dataset. However, there is considerable value in using that dataset in combination with the other two datasets.
First, the Decahose dataset provides a random 10\% sample of polls, which is invaluable for estimating the total number of polls published on Twitter, so we use it to address related questions RQ1 and RQ3. Decahose also offers many more samples than our query datasets, which is important when making precise temporal comparisons, so we use it together with the 2020 query dataset to address RQ2.3.
Second, the 2016 query dataset allows us to study the growth of social media polls (RQ1) by comparing it with the 2020 query dataset, as we do next.

\begin{figure}[t!]
    \centering
    \includegraphics[width=0.85\linewidth]{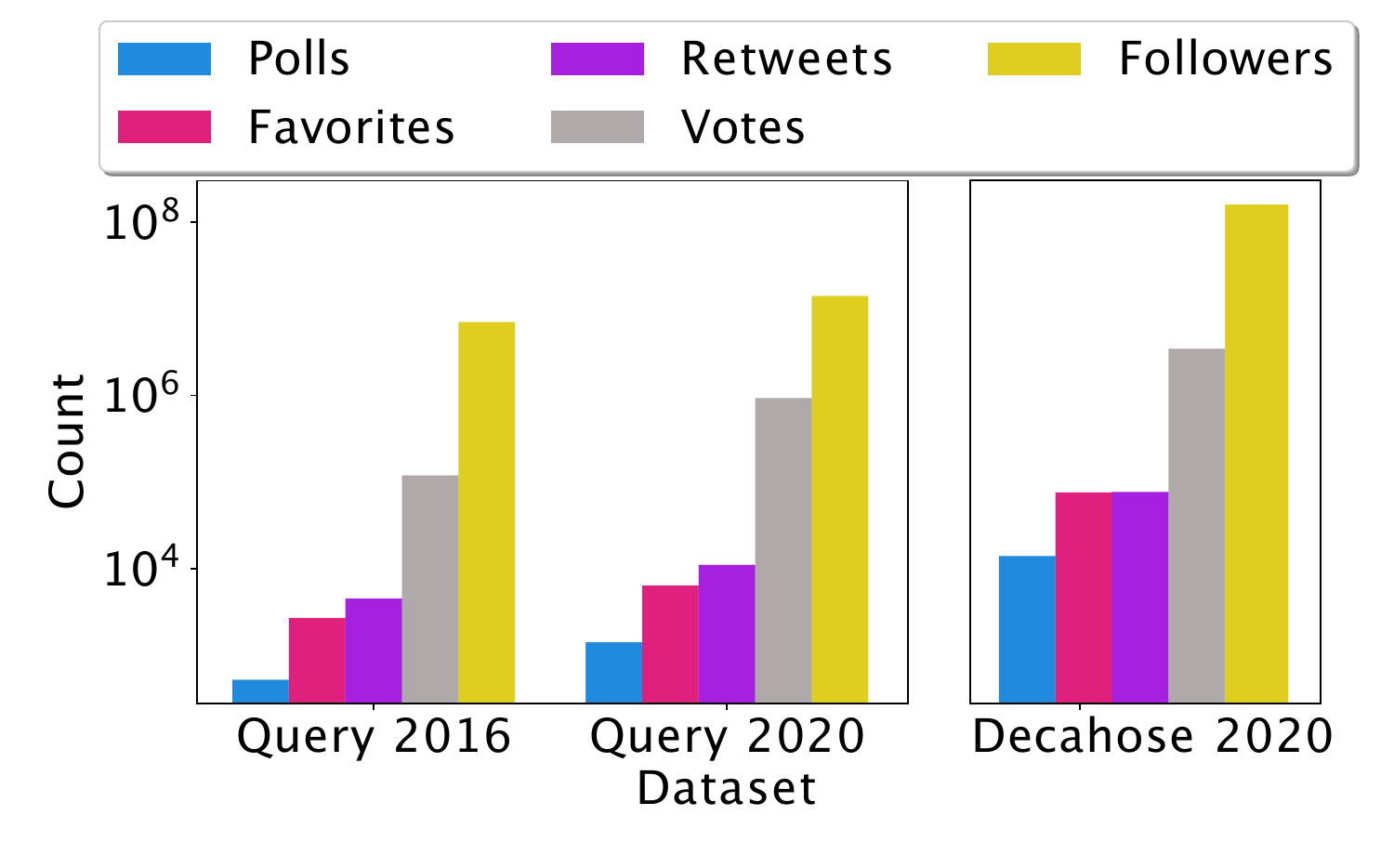}
    \caption{Growth of the number of polls, votes, poll retweets, poll favoriters, and followers of the poll authors. The difference between the numbers for 2016 and 2020 query datasets (left and center) illustrates the growth of polls. The Decahose dataset (right) includes only about 10\% of all polls, so the respective numbers represent \textit{only} about one-tenth of all election polls posted on Twitter. 
    The numbers suggest that election polls are becoming widespread.}
    \label{fig:datasets}
\end{figure}

\begin{figure}[t!]
	\centering
    \includegraphics[width=0.6\linewidth]{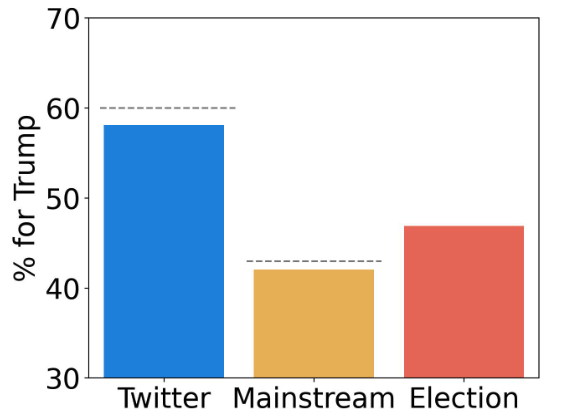}
	\caption{Average outcome of Twitter polls, 
 the average outcome of mainstream polls, and the official election outcome. Dashed lines signify the median.}
    \label{fig:bar-plot}
\end{figure}

\section{Rise of Election Polls on Twitter (RQ1)}
\label{sec:poll-growth}

We use our three large datasets of Twitter polls related to the 2016 and 2020 U.S. presidential elections to characterize the prevalence and growth of election polls on Twitter.
We plot the number of polls, votes, poll retweets, poll favoriters, and the number of followers of the poll authors for 2016 and 2020 for each of our datasets in Figure~\ref{fig:datasets}.
First, we note that all of the numbers have considerably grown between 2016 and 2020 (compare left and center of Figure~\ref{fig:datasets}), suggesting that election polls have grown in popularity on Twitter. 
Furthermore, the numbers shown in the figure underestimate the true numbers of polls and users engaging with them by a factor of 10 or more, since the Decahose dataset includes random 10\% of all polls, while the query datasets include less than that. 
Thus, the numbers of Decahose polls in Figure~\ref{fig:datasets} and Table~\ref{tab:datasets} represent \textit{only} about \textit{one-tenth} of all election polls posted on Twitter. For instance, there were nearly 13,000 polls in Decahose in 2020. 
Knowing that they constitute 10\% of all polls,\footnote{We confirm that the Decahose dataset contains 10\% of all polls published on Twitter, by counting the number of polls among all tweets published on September 20, 2022, based on a tweet dataset collected by a prior work \cite{pfeffer2023just}, and comparing it with the number of polls in Decahose.} we estimate that overall about 130,000 polls related to the U.S. presidential candidates were posted on Twitter in 2020, and they attracted over 20 million votes. 
This vote count does not include deleted or suspended polls.\footnote{If a poll is deleted or suspended after daily Decahose collection, then it is included in the Decahose. However, poll outcomes are not included in the Decahose and it is no longer possible to determine vote count and outcome for polls that have been deleted or suspended.}
For reference, there were about 168.42 million registered voters in the 2020 U.S. presidential elections. Finally, if the growth trend between 2016 and 2020 extends into 2024, we expect the number of polls and votes will be even larger in 2024.

\section{Biases in Results of Election Polls (RQ2.1)}
\label{sec:twitter-vs-mainstream}

\begin{figure*}[t!]
    \centering
    \includegraphics[width=0.85\linewidth]{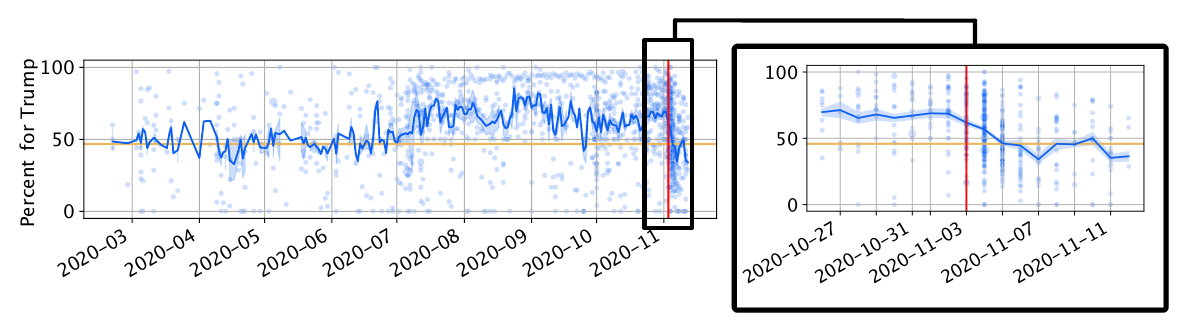}
    \caption{
    (Left) Moving average of the percent of votes for Donald Trump in Twitter polls published over the course of 2020. (Right) Zooming into the period of 3 weeks around the election days shows that the bias towards Trump diminished right after the election days, marked with the vertical red bars. The election outcome (orange line) is shown for comparison.}
    \label{fig:perc_for_trump_temp}
\end{figure*}

We compare the results of Twitter and mainstream polls gauging support for the 2020 U.S. presidential candidates.
Overall, the results of Twitter polls demonstrate that there is a substantial partisan slant, with a consistent leaning toward Trump (Figure~\ref{fig:bar-plot}). 
Specifically, the median support for Trump in social polls is 60\% (average 58\%), while in mainstream polls it is 43\% (average 42\%). 
While social media polls overestimate support for Trump compared to actual vote shares on election day, mainstream polls tend to underestimate Trump's support. 
We observe similar gaps in the 2016 election. 
We also note that the variance of Twitter polls is substantially larger than that of mainstream surveying (10 and 750, respectively), suggesting that social poll outcomes are affected by a number of diverse factors. 

Furthermore, we observe a significant drop in support for Trump immediately after the election days (see the right panel in Figure~\ref{fig:perc_for_trump_temp}). Several potential factors might contribute to this decline, such as digital astroturfing, differences in engaged online behavior between Trump and Democratic party supporters that wane post-election, or the influence of live election reporting. These findings shed light on the complex dynamics shaping social poll results and highlight the need for nuanced interpretation.

Partisan behavior may indeed differ between authors of Trump and Biden polls; in fact, there is reason to believe that Trump's coalition feels stronger about their support than that of Biden. 2020 election polling measures a 20\% gap in \textit{strong} support to Trump's favor and, as a result, the latter's voters may be more likely to post tweets, media, and polls related to his candidacy \cite{pew2020elect}.  Political analysis on this front is out of scope for this paper, and instead, we seek to determine to what extent other factors might feasibly explain the gap. We turn to these questions next.

\section{Questionable Votes in Social Polls (RQ2.2)} 
\label{sec:questionable-votes}

On June 19th, 2020, Polish state media \textit{TVP INFO} ran an article detailing a striking abnormality in a poll it published on its Twitter account asking who won a Polish presidential debate~\cite{tvp2020info}.
\textit{TVP INFO} claimed that 19,539 out of 35,202 votes (44.5\%) had been fraudulent, citing the 44.5\% difference between the private vote count reported in the tweet's analytics and the public vote count visible to the public. 

In this section, we corroborate the respective indicator of inorganic user behavior in Twitter polls: the discrepancy between the count of votes shown in public and the count shown privately to the poll author. 
We first verify that this discrepancy is ascribable to votes purchased on markets for inorganic user behavior. 
Then, we survey 984 authors (1,440 polls) from the 2020 query dataset for their private vote counts and estimate the fraction of questionable votes in the 2020 election polls as the discrepancy between public and private vote counts.

\subsection{Vote Count Discrepancy and Inauthentic Votes} 

\subsubsection{Methods} 

To emulate a potential agent seeking to distort an online poll, we searched on \textit{Google} for the query ``\textit{buy Twitter poll votes}.''
Next, we randomly picked five of the top 10 online stores (see full list in Appendix B). Then, for each of the five stores, we posted two identical polls and purchased 100 fake votes from the store per poll. Using a separate account, we submitted one organic vote for the opposite candidate choice than the inauthentic votes.
We ran the polls for 24 hours, then tabulated and compared the publicly-listed and privately-listed vote counts found in the user analytics (Appendix C).

\begin{table}[h!]
\centering
\resizebox{\columnwidth}{!}{
\begin{tabular}{c c c c c c} 
\toprule
 & & \multicolumn{2}{c}{True vote counts} & \multicolumn{2}{c}{Twitter counts}\\
 Trial & Vendor of votes & Organic & Bought & Public & Priv. \\ 
\midrule
 1 &    viplikes.net & 1 & 100 & 138 & 1 \\
 2 & viplikes.net & 1 & 100 & 120 & 1 \\
 3 & socialboss.com & 1 & 100 & 134 & 1 \\
 4 & socialboss.org & 1 & 100 & 125 & 1\\
 5 & socialwick.com & 1 & 100 & 135 & 1 \\ 
 6 & socialwick.com & 1 & 100 & 126 & 1 \\ 
 7 & gettwitterretweet.com & 1 & 100 & 124 & 1 \\
 8 & gettwitterretweet.com & 1 & 100 & 120 & 1 \\
 9 & instafollowers.com & 1 & 100 & 120 & 1 \\ 
\bottomrule
\end{tabular}
}
\caption{Discrepancy between public and private vote counts.  Fake votes purchased in the experiment are included in the public vote count, but not in the private vote count.}
\label{tab:paidvotes}
\end{table}

\subsubsection{Results}

We study whether public vote counts differ from private vote counts by the margin of fraudulent votes. 
Table~\ref{tab:paidvotes} shows that---regardless of cause---Twitter's poll system records public votes that its private analytics does not acknowledge. Although only 100 votes were purchased in every case, vendors consistently provided more votes than that (between 120 and 140 votes), which likely corresponds to an attempt to avoid automatic detection. The private count of votes shows the correct number of organic votes provided on the polls, that is, one organic vote. In other words, none of the purchased votes counted towards the private count. We list potential explanations in the Discussion and Conclusions section.

\subsection{Vote Count Discrepancy in Election Polls}

\subsubsection{Methods} To estimate the number of questionable votes in our tweet poll dataset, we contacted 984 authors from the 2020 query dataset and requested the private vote count for their corresponding poll(s). 
It is no longer feasible to check private vote counts for older posts due to the changes to the Twitter user interface at the end of 2022. Hence, we could not conduct the same survey for the 2016 polls. Our survey and analysis of responses received IRB approval.

We sent each user a direct message or a tweet introducing the researchers,\footnote{See Appendix A for full message contents} before soliciting the private vote count of their respective poll(s). The message contained a link to the poll in question and instructions on retrieving the private vote count. It also clarified that the researchers would maintain the author's anonymity and only publish aggregate results. 

To better interpret these results, we performed a placebo experiment in which we posted a Twitter poll and solicited responses from our colleagues working at the college of the last author, by sending an email to an internal mailing list, without explaining the specific purpose of this experiment. To the best of our knowledge, this poll contained human responses only, without the presence of inauthentic votes. We leverage the public/private vote count discrepancy measured in this placebo experiment to contextualize those of the surveyed poll authors.

\subsubsection{Results}
Only 22 poll authors responded to our request, despite our multiple attempts to contact the users who did not respond, probably because of the sensitive nature of the request, e.g., some of the users might have participated in political campaigns. While the number of responses is not high, they offer a significant insight into political polling on Twitter consistent with the other parts of this study.

Of the 22 poll authors, 21 users reported discordance between public and private vote records, with a median 35\% increase from private to public. We refer to this discrepancy as \textit{questionable votes}. Across all 22 polls, the total count of public votes is 3605, while the private vote count is 1555. Thus, 57\% of all votes appear questionable. The fraction of questionable votes among different polls varies drastically, from 0\% to 68\%. The outcomes of these polls are mixed, supporting 60\% for Trump on average, matching the overall average of 58\% for our 2020 dataset. 

For comparison, our placebo experiment, in which no inorganic votes were placed, demonstrated 40\% of its votes as questionable (33 out of 82). 
Thus, some discrepancies between the public and private vote count may be normal and expected, e.g., Twitter's classification of inauthentic votes may be inaccurate and always result in a fraction of votes classified as ``questionable''. However, even if we assume that $40\%$ of public votes (as determined by the placebo experiment) are not counted as private votes due to legitimate reasons, we still see potential evidence of manipulation.

\begin{figure}
    \centering
    \begin{picture}(\linewidth,80)
    \put(0,0){\includegraphics[width=1\linewidth]{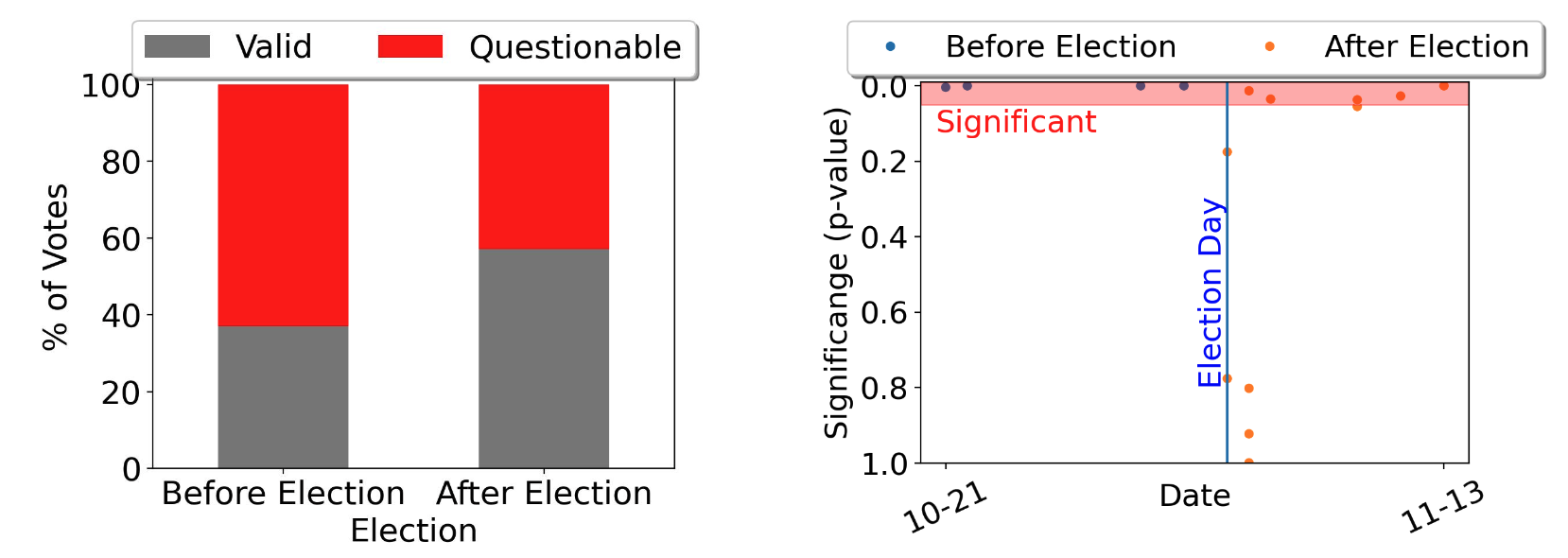}}
    \put(95,60){\small{A}}
    \put(215,60){\small{B}}
    \end{picture}
    \caption{The fraction of questionable votes drops---comes back to the expected fraction---\textit{after} the election day (left). The fraction of questionable votes is larger than expected for all evaluated polls before the election day and for a subset of polls after the election (right).}
    \label{fig:pub/priv-time}
\end{figure}

First, the fraction of questionable votes is significantly higher ($p=0.05$, Mann-Whitney U test) before than after the election day (Figure \ref{fig:pub/priv-time}A). 
Second, for each poll, we test whether the fraction of questionable votes is significantly higher than the $40\%$ of votes that may originate from legitimate sources.
We conduct a binomial test comparing the observed and expected number of private votes, given that $p=0.4$ and $n$ is the public vote count. We find that 13 of the 21 polls exhibit a larger discrepancy between the public and private vote count than expected by this binomial model ($p=0.05$, see Figure \ref{fig:pub/priv-time}B).
Furthermore, all polls posted \textit{before} the election have significantly more questionable votes than we would expect under the assumption that $40\%$ of public votes are not counted as private votes due to legitimate reasons.

\section{Changes in Suspicious User Activity (RQ2.3)}

In addition to scrutinizing polls, we analyze poll authors and participants (favoriters and retweeters; the identity of voters is not disclosed by Twitter) for signals of inorganic activity. We determine whether users are bot-like, foreign, or hyperactive accounts. While bot scores have an intuitive relationship to manipulation, signals of foreign origin (non-U.S) and hyperactivity may be markers of coordinated astroturfing campaigns, foreign or otherwise. Here, we analyze the query and Decahose sets of 2020 election polls jointly.

\subsubsection{Reference set of polls matched to election polls}
In this section, to identify suspicious user activity, we compare bot scores, locations, and activity levels of users participating in election polls with those participating in random polls. To this end, we collected a set of random polls from 2020, including non-political polls. Then, for each election poll from our two 2020 datasets, we match a poll to one with a similar vote count from that set of random polls. Finally, for each matched poll, we collect all its retweeters and favoriters to compare them with retweeters and favoriters of election polls.
We refer to that matched poll set as \textit{reference} polls.

\subsection{Bots}

\begin{figure}
    \centering
    \begin{picture}(\linewidth,100)
    \put(0,0){
    \includegraphics[width=1.03\linewidth]{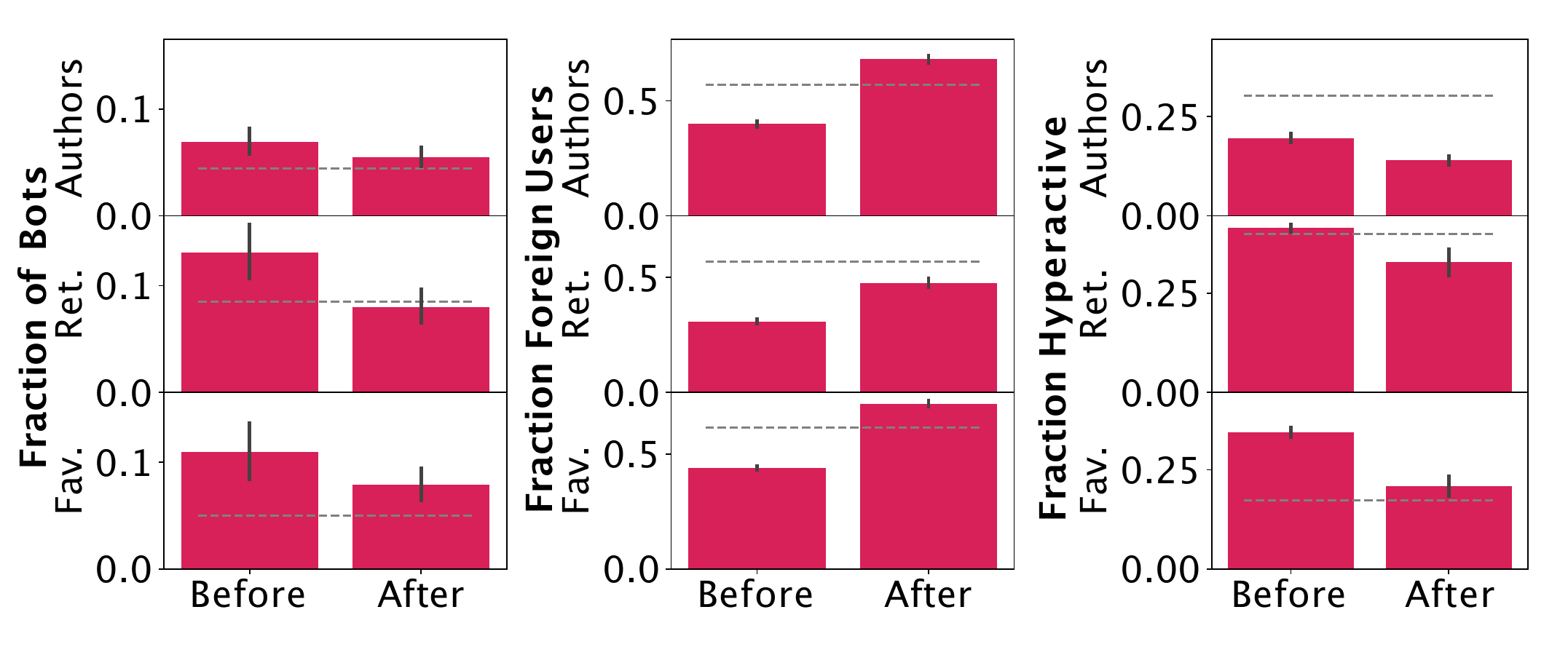}}
    \end{picture}
    \caption{The fraction of (from left to right) bots, foreign accounts, and hyperactive users among (from top to bottom) poll authors, retweeters, favoriters before and after the election day (left and right bars within each panel). The dashed line corresponds to the respective fraction for random reference polls matched to the election polls based on the number of votes and publication year.}
    \label{fig:before-after}
\end{figure}

\subsubsection{Methods} We subjected authors, retweeters, and favoriters of 2020 election polls through the machine learning classifier \textit{Botometer-V4} (MIT license) to estimate the distribution of human and bot accounts. \textit{Botometer} is a random forest classifier that evaluates network, user, friend, temporal, content, and sentiment features to label a profile as authentic or artificial \cite{sayyadiharikandeh2020detection}. To contextualize these results, we compared the outputs to a random sample of Twitter polls posted over the same period and of similar vote distribution. We then split user groups temporally--two weeks before and after the election--to identify elevated bot presence.

\subsubsection{Results} 
Figure \ref{fig:before-after} visualizes the percentage of bots among poll authors before and after the election in comparison to a random sample. The distributions for retweeters and favoriters follow the same qualitative patterns. Overall, we find that users participating in election polls have higher bot scores than users participating in the reference polls. Bot score distributions of poll authors differ significantly from our random sample ($ p < 10^{-10} $, MWU test), and significance also holds for retweeters and favoriters.~\footnote{See Appendix D for all significance values relevant to RQ2.3.} Further, we find that the fraction of bots among retweeters and favoriters of political polls posted two weeks prior to election day are significantly higher than those two weeks afterward ($ p <10^{-5} $ and $ p < 10^{-10}$, respectively; MWU tests).

\subsection{Foreign Accounts}

\subsubsection{Methods} 
On Twitter, users can optionally disclose their location in plain text using the \texttt{location} field of their profile. We resolved such entries to geolocations using Photon, an open-source geocoder built for OpenStreetMap data under the Open Database License.\footnote{\url{https://photon.komoot.io/}} 
To expand coverage, for users whose location could not be geocoded via the previous method, we combined the \texttt{location} and \texttt{description} plain-text fields of user profiles and extracted emojis corresponding to national flags, excluding the cases of users displaying flags of multiple countries. 
Based on this information, we distinguish between users belonging to foreign countries and the U.S. 

\subsubsection{Results} Similar to bot accounts, the distributions of foreign accounts for authors, retweeters, and favoriters differ significantly from the respective random samples. Conversely, however, occurrences of foreign accounts are \textit{less} likely in political polls. Additionally, the prevalence of foreign authors is significantly higher ($ p < 10^{-10}$, MWU test) post-election in comparison to pre-election. 

\subsection{Hyperactive Users}

\subsubsection{Methods} We refer to users who post, on average, more than 20 tweets per day as hyperactive. For all authors, retweeters, and favoriters, we estimate user activity as the average number of statuses (including retweets) since the creation date of the account. 

\subsubsection{Results} Activity distributions for authors, retweeters, and favoriters differ significantly from their respective reference ($p < 10^{-10}$ for all groups, MWU tests). Indeed, across the three groups, activity before the election is higher than post-election ($p < 10^{-10}, p = 3.8 \times 10^{-8}, p < 10^{-10}$ MWU tests, respectively). Figure \ref{fig:before-after} visualizes the percentage of hyperactive users before and after the election in comparison to a random sample. 

\section{Traits of Users Engaging with Polls (RQ2.4)}
\label{sec:user-attributes}

The biased outcomes of polls may be explained by factors besides suspicious activity in election polls, e.g., a skewed demographic and political composition of users involved in polling on Twitter.
Here, we study the characteristics of users who author and vote in polls in the 2020 Query dataset (see Table \ref{tab:datasets}). 
Since Twitter does not reveal the identity of poll voters, we consider poll retweeters and favoriters as likely voters. 
We study how the three user groups deviate from a representative sample of the U.S. population in terms of their age, gender (in comparison to the U.S. Census), and political ideology (in comparison to the 2020 U.S. presidential election exit polls).  
To quantify each of these user characteristics we apply state-of-the-art user attribute inference methods, described next.

\subsection{Age and Gender}

\subsubsection{Methods} 
We classify the gender and age of users, attributes commonly understood to correlate with voting behavior. To do so, we employ the multilingual, multimodal, and multi-label machine learning tool \textit{M3-Inference} under MIT license \citep{wang2019inference}. \textit{M3-Inference} is a deep learning text and image model that uses usernames, profiles, and photos to infer age and gender with state-of-the-art accuracy while diminishing algorithmic bias in comparison to other approaches.
Since the model additionally infers the likelihood that the given account represents an organization, we exclude from our analysis those users who exceed an org score of $0.90$.

\subsubsection{Results} 
\begin{figure}[t!]
	\centering
	\includegraphics[width=0.9\columnwidth]{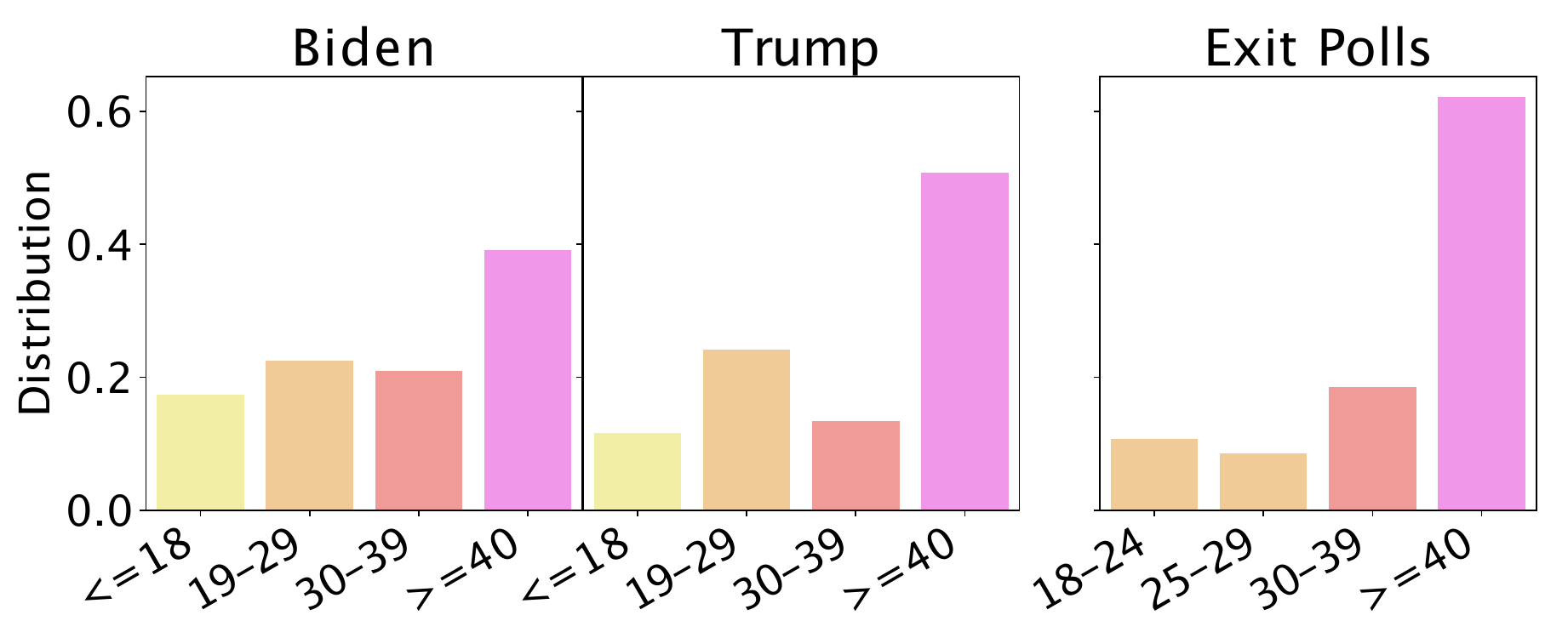}
	\caption{Age distribution of authors of 2020 U.S. election polls on Twitter. For comparison, the rightmost figure shows gender distribution for the 2020 exit polls. The bars are color-coded to mark the correspondence between the age brackets, e.g., the second bin for social polls corresponds to the first two age bins for exit polls.}
	\label{fig:demo-age}
\end{figure}

First, we find that the fraction of males is about 2 times larger among poll authors than among exit poll respondents (76\% vs. 49\%) and that the fraction of poll authors below 30 years old is almost 3 times larger (Figure~\ref{fig:demo-age}). Our results conform to prior research suggesting that Twitter skews heavily male and young \citep{mellon2017population}. Trump polls, i.e., polls won by Trump, lean more male than Biden polls (83 \% vs 77.1\%), which corresponds to his voting base.  
These biases are greater among authors and their followers (58\%) than among retweeters (53\%) and favoriters (52\%), suggesting that while young males are mobilized to author polls, people engaging with the polls are more similar to the general population in terms of age and gender. Retweeters and favoriters are comparably older (and more representative of U.S. voters) than authors and followers (42\% and 50\% have age $\geq 40$ vs 30\% and 32\%, respectively).

\begin{figure}[t!]
	\centering
	\includegraphics[width=0.9\columnwidth]{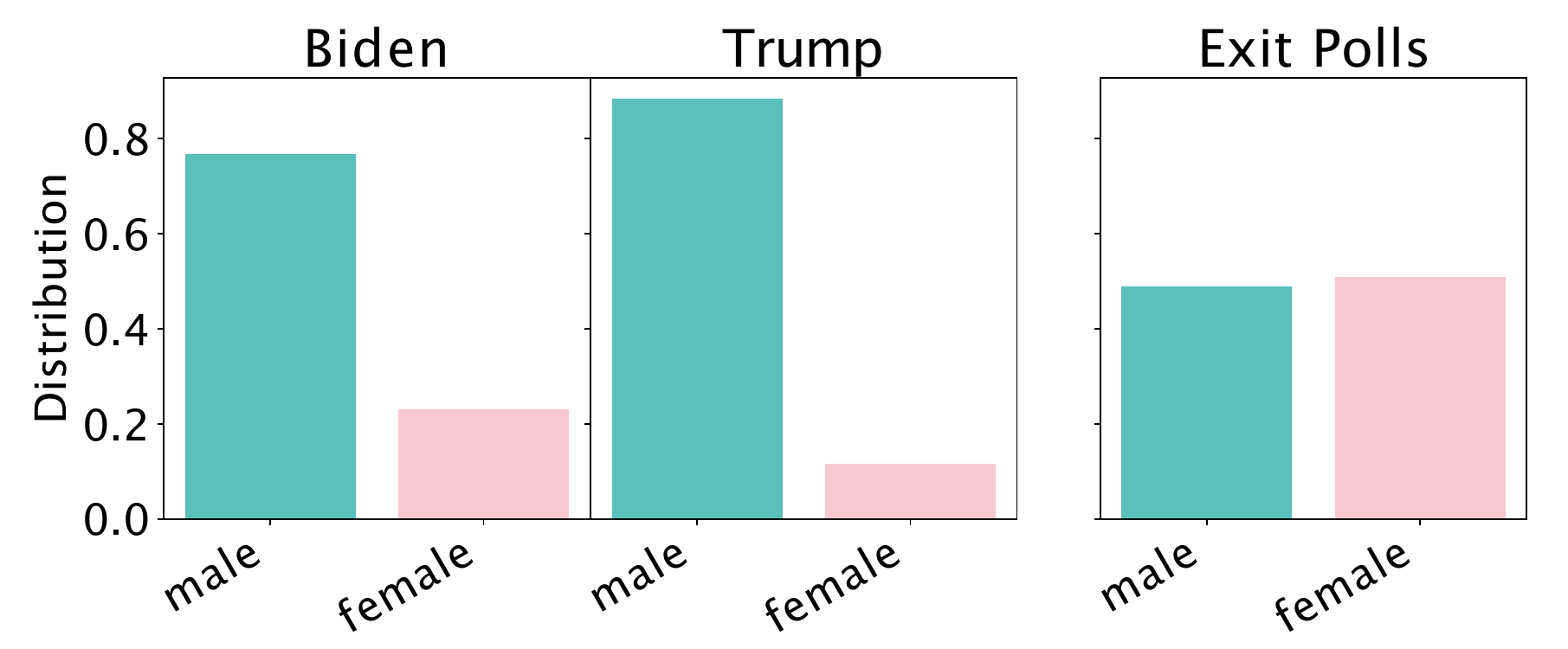}
	\caption{Gender distribution of authors of 2020 U.S. election polls on Twitter. For comparison, the rightmost figure shows gender distribution for the 2020 exit polls.}
	\label{fig:demo-gender}
\end{figure}

\subsection{Political Ideology}

\begin{figure}[tb]
    \centering
    \includegraphics[width=1\linewidth]{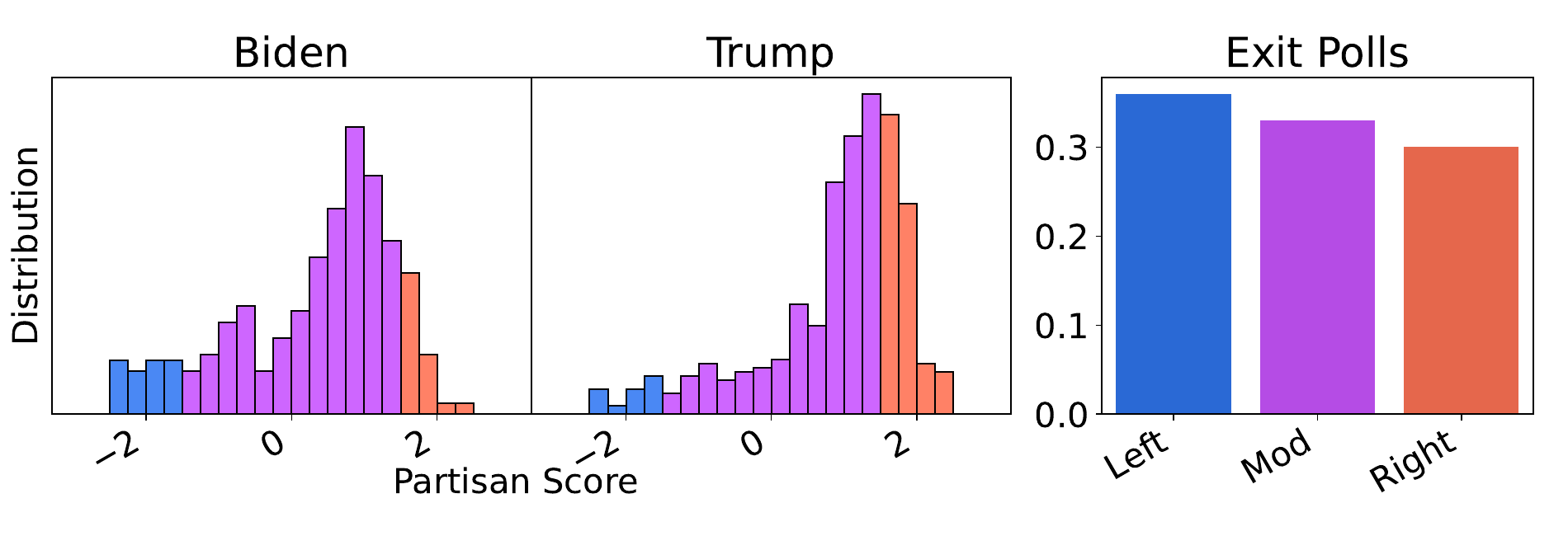}
    \caption{Distribution of partisanship scores for poll authors for social media polls, separately for Biden and Trump (left and center), and exit polls (right).}
    \label{fig:partisanship-authors}
\end{figure}

\subsubsection{Methods}
We estimate relative political ideology in Twitter polls using a Markov chain Monte-Carlo approach by the authors of \citep{barbera2015tweeting, Barbera2015} under MIT license. 
The tool infers the political ideology of a user based on a set of users who that user follows (so-called \textit{followees}); that is, if user $A$ in the majority follows well-known right-wing accounts and user $B$ follows well-known left-wing accounts, the tool will output a positive value for user $A$ and a negative value for user $B$. 
Each user instance is mapped to a continuous political ideology value in the interval $[-3, 3]$.
While we provide the raw distribution of these values, in the next section (RQ2.5) we discretize this range into three bins (\textit{Left}, \textit{Moderate}, \textit{Right}), splitting the space evenly for simplicity. Prior work by \citep{barbera2015tweeting} shows that this approach performs comparably with standard ideological assessment surveys. 
The model infers users' political affiliation from their followees---the users an account is following. Due to rate limitations of the Twitter API and the removal of Academic API access, collecting the followees for Decahose users is infeasible. 

\subsubsection{Results} 

To compare Twitter users' political ideology scores with the political affiliation data from exit polls, we converted the continuous scale of inferred political ideology of users into a discrete scale. Also, for the purposes of this discussion, we use the terms political ideology and political affiliation interchangeably. 
Our analysis shows that the distributions of political ideology of poll authors is skewed towards the right (Figure \ref{fig:partisanship-authors}). This result is consistent with the fact that social poll results are skewed towards Trump (RQ2.1). Interestingly, retweeters and favoriters of social polls are even more likely to be more conservative than the poll authors themselves (figure not shown). This asymmetry resembles the one observed in the distribution of political ideology of users who interact with misinformation, which also predominantly leans to the right~\citep{nikolov2021right, gonzalezbailon2023asymmetric}.

\section{Correlates of Bias in Poll Outcomes (RQ2.5)} 

\label{sec:regressions}
The previous sections painted a broad picture of the sources of bias that are potentially present in Twitter election polls. Next, we relate such biases to poll results, to assess the relative strength of their relation with poll outcomes. 

\subsubsection{Methods} 

We analyze polls from the 2020 query dataset.
We perform ordinary least squares regression to relate potential sources of bias to poll outcomes, operationalized as the fraction of votes cast in favor of Trump. Building on the observations from the previous sections, we consider potential sources of bias associated with the characteristics of both the authors of the polls and their respondents; like in the previous section, we use retweeters and favoriters as proxies for respondents. 
We include as independent variables several sociodemographic attributes: the users' gender (male or \textit{female})\footnote{We indicate in italics the reference levels of the variables.}, age group (\textit{less than 30}, between 30 and 39, greater than 40), political ideology (democrat, \textit{moderate}, republican), and location (U.S. red state, U.S. blue state, \textit{U.S. swing state})\footnote{We source the colors of the US states from the list available on Wikipedia, based on the outcomes of the 2016 election \url{https://en.wikipedia.org/wiki/Red_states_and_blue_states}}. 
Furthermore, we include three additional user attributes: foreign location to the U.S. (yes or \textit{no}), bot score (bot or \textit{not bot}, using the threshold 0.83 for bot users), and hyperactivity (yes or \textit{no}, using the threshold of 20 tweets/day for hyperactive users). Past work related such attributes to information operations, and as such they may relate to manipulation attempts.\footnote{We caution the reader that such attributes do not imply malicious behavior: e.g., political campaigns that schedule multiple tweets a day would likely be considered both bots and hyperactive, although their use of the platform is legitimate.} We encode user traits as categorical for poll authors and as probability distributions of traits of retweeters or favoriters of a poll. 
To reduce noise in the dependent variable and the number of missing independent variables, we exclude polls with fewer than $M=60$ votes. We choose $M$ by maximizing $R^2$ and normalized Akaike information criterion~\cite{cohen2021normalized}.
We impute missing values in the remaining polls ($N=595$) substituting them with the mean of the corresponding independent variable.

\begin{table}[t!]
\begin{center}
\resizebox{0.85\columnwidth}{!}{
\begin{tabular}{lrl}
\toprule
\textbf{Independent variable} & \textbf{coef} & \textbf{P$> |$t$|$}\\
\midrule
{const}                                      &       0.42  &    ***\\
{$p_{a}$(gender=male)}            &       0.01  &    \\
{$p_{r}$(gender=male)}                  &      -0.01  &    \\
{$p_{f}$(gender=male)}             &      -0.03  &    \\
{$p_{a}$(age$ \in[30,39] $)}              &      -0.01  &    \\
{$p_{r}$(age$ \in[30,39]) $}                    &      -0.01  &    \\
{$p_{f}$(age$ \in[30,39]) $}               &       0.00  &    \\
{$p_{a}$(age$ \ge 40$)}             &      -0.01  &    \\
{$p_{r}$(age$ \ge 40$)}                   &       0.11  &    **\\
{$p_{f}$(age$ \ge 40$)}              &       0.04  &    \\
{$p_{a}$(ideology=dem)}           &       0.00  &    \\
{$p_{r}$(ideology=dem)}                 &      -0.15  &    *\\
{$p_{f}$(ideology=dem)}                        &      -0.12  &    *\\
{$p_{a}$(ideology=rep)}            &       0.12  &    ***\\
{$p_{r}$(ideology=rep)}                  &      -0.03  &    \\
{$p_{f}$(ideology=rep)}                        &       0.28  &    ***\\
{$p_{a}$(location=blue state)}                           &       0.03  &    \\
{$p_{r}$(location=blue state)}                                 &      -0.04  &    \\
{$p_{f}$(location=blue state)}                            &      -0.03  &    \\
{$p_{a}$(location=red state)}                            &       0.04  &    \\
{$p_{r}$(location=red state)}                                  &      -0.00  &    \\
{$p_{f}$(location=red state)}                             &      -0.01  &    \\
{$p_{a}$(location=foreign)}              &      -0.03  &    \\
{$p_{r}$(location=foreign)}                    &      -0.05  &    \\
{$p_{f}$(location=foreign)}               &      -0.09  &    **\\
{$p_{a}$(bot=yes)}                            &       0.22  &    **\\
{$p_{r}$(bot=yes)}                                  &       0.01  &    \\
{$p_{f}$(bot=yes)}                             &       0.02  &    \\
{$p_{a}$(hyperactive=yes)} &       0.09  &    ***\\
{$p_{r}$(hyperactive=yes)} &      -0.08  &    \\
{$p_{f}$(hyperactive=yes)}  &       0.07  &    \\
\midrule
\textbf{Dependent variable:} & \% for Trump&\\
\textbf{{No. observations:}}                          &         595&\\
\textbf{Adj. $\mathbf{R}^2$}:     &     0.514&\\
\bottomrule
\end{tabular}
}
\end{center}
\caption{Parameters for the ordinary least squares regression model using percent support for Trump as dependent variable. Coefficients with subscript \textit{a}, \textit{r}, \textit{f} refer respectively to poll authors, retweeters, and favoriters. We indicate statistical significance at levels $p<0.001$ (***), $p<0.01$ (**), and $p<0.05$ (*).}
\label{tab:regression}
\end{table}

\subsubsection{Results} The model explains a large fraction of the variance of poll outcomes, with an adjusted $R^2=0.514$. Thus, we further the analysis and interpret the coefficients associated with the independent variables. The model shows support for most of our hypotheses. Higher rates of support for Trump are expressed in polls with older demographics (the coefficients are statistically significant for retweeters), as well as  users leaning Republican (authors and favorites). Conversely, higher support for Biden is associated with users leaning Democrat (retweeters and followers). We also find that users located in foreign countries (favoriters) are associated with polls that support Biden; this finding is in line with accounts of support for Biden and disfavor for Trump in NATO countries \cite{howorth2021europe, sintes2022europe}. Furthermore, the model shows a higher likelihood of the presence of automated and hyperactive accounts among the authors of the polls that support Trump, which echoes previous findings about the heightened activity and strategic use of Twitter as a political medium by Trump supporters. 
Contrary to our intuition, we do not observe a connection between support for Trump and users' gender, despite a larger fraction of males in polls won by Trump (Figure~\ref{fig:demo-gender}); nor do we observe differences between users located in historically red or blue states: these associations are statistically insignificant when controlling the remaining variables in the model.

The quality of the model fit suggests that it may be possible to infer poll biases from the characteristics of the user base. However, the fact that significant coefficients in the model are associated with different, specific user groups asserts the importance of properly accounting for the different roles that users play in the generation of polls, their promotion, and engagement with them. Note that all independent variables are stable traits of the users that can be known in advance of the polls' outcomes, which makes them a promising avenue for predictive applications such as the post-stratification of poll outcomes to address their biases and the use of polls for the social sensing of public opinion.

\section{Polls Spreading Voter Fraud Beliefs (RQ3)}
\label{sec:conspiratorial}

Here, we identify polls that express skepticism towards traditional media's coverage of the elections, distrust poll results from mainstream sources, or question the legitimacy of the electoral process. We call such polls ``conspiratorial'', as they often expressly support popular conspiracy theories about the elections. We estimate their overall number and the level of user engagement with the polls.

\subsubsection{Methods} 
To identify conspiratorial polls, we analyze the Decahose polls using election-related keywords. We take a human-centered approach as follows. First, two authors manually evaluated a random sample of 1,000 polls to code whether the question wordings or response options were conspiratorial. This initial coding round identified 19 conspiratorial polls (coder 1: N = 19, coder 2: N = 23) with high agreement between annotators (Krippendorff's $\alpha$ = 0.76). The two coders resolved disagreements through discussion with a third coder for the final annotation of this sample.

\subsubsection{Results} 

Knowing that there are about 19 conspiratorial polls among 1,000 random election polls, 
and building on the results of RQ1 that estimated a total of 130,000 election poll,
we estimate that about 19*130=2,470 conspiratorial polls were posted on Twitter in 2020. 

\setlength{\columnsep}{14pt}
\begin{wraptable}[16]{r}{0.45\columnwidth}
    \centering
    \begin{tabular}{c|c}
    \toprule
    \multicolumn{2}{c}{85 conspiratorial polls} \\
    \toprule
    Retweets & 541\\
    Favorites & 424\\
    Votes  & 10,075\\
    Followers & 997,025\\
    \bottomrule
    \end{tabular}
    \caption{The numbers of retweets, favorites, and followers of the identified 85 conspiratorial polls. The vote count is the total for the 47 (out of the 85) conspiratorial polls that remain available on Twitter.}
    \label{tab:conspiratorial}
\end{wraptable}

Thus, our set of 85 conspiratorial polls is a very small subset (about 3.4\%) of the total: note that although the dictionary-based approach is straightforward, it underestimates the number because some conspiratorial polls do not include the keywords or do not appear in Decahose. Nonetheless, the extended sample of conspiratorial polls sheds light on the reach of conspiratorial polls on social media (Table~\ref{tab:conspiratorial}), their use, and on the efforts to moderate such polls. 
Most of these conspiratorial polls were published after the presidential election day.
Of the 85 identified conspiratorial polls, 38 were unavailable at the time of writing due to deletions (29 polls) and account suspensions (9 polls). This suggests that either poll authors removed their polls or there was a moderation effort on the platform to curb conspiratorial information.

Through qualitative coding, we unpack the themes in the content of the sample. Conspiratorial polls predominantly reflect skepticism toward the accuracy and neutrality of mainstream media polls. Many of them express a lack of trust in polls conducted by major news outlets and national polling organizations, with a recurring theme of challenging these polls' results through independent, user-conducted polls on social media platforms. There is a notable emphasis on achieving ``unbiased'' or ``accurate'' polls, indicating a perception that existing polls might be biased, especially toward Trump. The wording of the questions in some polls also brings up issues related to election fraud and voter suppression. This sample collectively underscores a mix of distrust in traditional polling methods, a desire for alternative, user-generated data, concerns about political bias and election integrity, and the contentiousness of the 2020 elections.

\section{Discussion and Conclusions}

According to our study, there were about 20 million votes cast in about 13,000 Twitter polls related to the 2020 presidential elections.
On average, these polls show 58\% support for Trump and 42\% for Biden, in a striking reversal of the true election result, in which Trump achieved 46.8\% of votes and Biden 51.3\%.
Future research can estimate the effect of such polls on the beliefs and voting intentions of people interacting with them. 
Especially, this study finds that about 1.9\% of polls on the 2020 elections questioned the validity of mainstream polls and/or election results. Further research is necessary to evaluate whether social polls can reinforce misinformed voter fraud beliefs.

Whenever platform transparency is lacking, researchers are forced to consider a breadth of potential explanations for otherwise opaque behavior. 
In the 2020 presidential election polls, we find signals of questionable votes, the activity of bots and hyperactive users, as well as demographic biases. 

In particular, we find that Twitter's poll system misreported public vote counts compared to what was shown to poll authors. 
In our experiments, these discrepancies neatly match purchased votes.
This result and the opacity of Twitter that enabled the discrepancy are troubling.
The difference in vote counts could be attributed to a bot classifier on Twitter's end, filtering out votes deemed suspicious. Considering that Twitter's business model incentivizes perceived activity and interest, the complete filtering-out of votes may prove unprofitable. Under this framework, Twitter's interests as a platform run parallel to those of online astroturfers and, in effect, against that of the public good, which makes transparency efforts all the more necessary. 

Our demographic analysis shifts the framing from \textit{disinformation} to \textit{misinformation}. The over-representation of certain strata of the overall population broadcasts a biased message, which contradicts and potentially breeds suspicion of institutional polling. We show that bias along age, gender, and partisan lines explains a large fraction of poll outcome variance. 

The opacity of Twitter polls makes estimating the effect of biases, either through misinformation or disinformation, difficult to estimate. Twitter's discontinuation of academic access programs further exacerbates the problem. Data access to researchers, ensured through policies such as the bipartisan Platform Accountability and Transparency Act proposed by U.S. legislators and the Digital Services Act in Europe, may be a first step towards a solution. However, we can also imagine action on Twitter's part (and other social media platforms, for that matter) to better inform its users about misleading biases in election polls. 

Finally, we recognize that the non-representativity of Twitter users engaging in polls may be turned into useful information and leveraged as complementary to traditional representative surveys, and used in combination with them: hard-to-reach populations, such as those resistant to institutionalized polling, may be eager to disclose their preferences to ideologically aligned social poll authors.

\subsection{Limitations}
The findings of this study may not generalize beyond the polls gauging support for the 2016 and 2020 U.S. presidential election candidates. We look forward to similar studies for more recent elections, including elections in other countries, particularly the ones experiencing significant voter fraud conspiracies. 

This study explains biases in poll outcomes with biases in characteristics of users who might have participated in these polls. However, this relationship may be solely correlational, since (i) user demographics and political ideology are inferred after respective polls were conducted, (ii) polls showing biased outcomes may encourage votes from users having certain characteristics.
Finally, further research is needed to establish a causal link between exposures to biased social poll outcomes and voter fraud beliefs.

{\fontsize{9pt}{10pt}
\selectfont
\bibliography{biblio}}

\vspace{3cm}

\section{Ethics and Competing Interests}

One crucial aspect of the design of this study is that it involved interacting with the authors of political polls through direct messaging. This enabled gaining unprecedented information on private vote counts, only visible to such users. 
After deliberation, the researchers decided not to disclose that the discrepancy may be the result of bought votes because the discrepancy was considered to be a phenomenon worth investigating regardless of this hypothesis. In particular, the poll authors may have colluded in the manipulation as well as have been in good faith. In either case and regardless of the truth of the hypothesis, the stigma associated with the presumption of collusion may have led to misreporting.

To gain evidence on the process of vote manipulation, the research design required purchasing specific numbers of votes by sellers of inauthentic user behavior. The authors took this matter seriously before determining its necessity for the study. The authors do not endorse such a market---in fact, the results of the present research aimed at exposing and ultimately reducing it, which was evaluated as a net positive when weighted against the small sum invested in the experiment. The authors took care of making the polls in the experiment invisible to regular Twitter users, so as not to expose the latter to manipulated results.

\section*{Appendix A}
Table \ref{message} contains the full message sent to poll authors.
\begin{table}[ht]
\centering
\begin{tabular}{|p{0.45\textwidth}|}
\hline
Hello! I'm a researcher at \textbf{REDACTED} studying Twitter polls. I've noticed that in some instances on Twitter, the site's public votes shown on user timelines differ from the private votes listed in Tweet Analytics, and I'm studying potential causes and frequency of this phenomenon. 

On \textbf{END\_DATE} you posted \textbf{LINK\_TO\_TWEET}. Would you be willing to anonymously contribute to our study by disclosing the number of recorded votes displayed within Tweet Analytics? Your privacy will be fully preserved: we will study the numbers of votes on aggregate across hundreds of polls without revealing their individual values. If you choose to participate, we will share with you the findings of our study. 

To participate in the study, please simply copy-paste the following number of votes as a response to this message. To see the number, please click on the above link and navigate to “View Tweet Analytics”. There should be an attribute labeled “Votes” under the "View all engagements" button. Please simply respond to this message by copy-pasting that number (of “Votes” in “View Tweet Analytics”). Please double-check that you’re copying the right number correctly. \\ \hline
\end{tabular}
\caption{Message sent to surveyed authors regarding obtaining private vote counts}
\label{message}
\end{table}

\section*{Appendix B}
The stores returned by Google are shown in Table~\ref{tab:services}.

\begin{table}[h!]
\centering
\begin{tabular}{c c} 
 \hline
 Rank & Name \\ 
 \hline\hline
 1 & viplikes.net  \\
 2 & socialwick.com \\
 3 & buytwitterpollvotes.com\\
 4 & socialboss.org \\
 5 & gettwitterretweet.com \\ 
 6 & famousfollower.com \\ 
 7 & socialyup.com \\ 
 8 & rousesocial.com \\  
 9 & instafollowers.com \\
 \hline
\end{tabular}
\caption{Top 10 Google ranking of ``\textit{buy Twitter poll votes}'' query retrieved on 12/5/2021}
\label{tab:services}
\end{table}

\section*{Appendix C}

\begin{figure}
    \centering
    \includegraphics[width=0.45\textwidth]{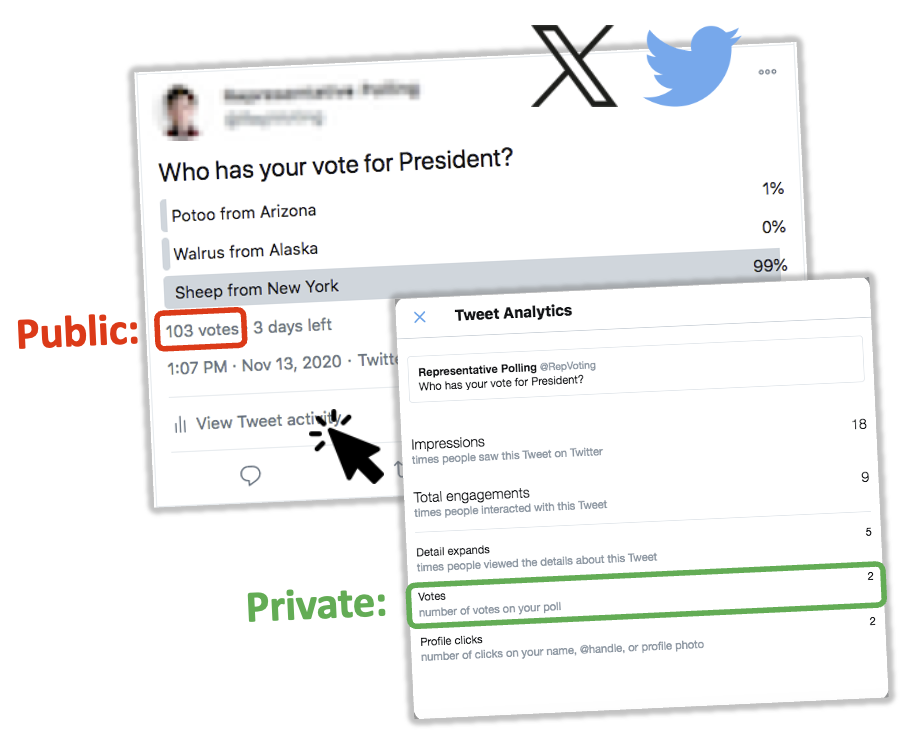}
    \caption{A poll used in the fake vote experiment. ``Private'' (blue box) vote counts are distinct from those listed publicly (red box) under the tweet and are visible to the poll author under the ``View Tweet Activity'' tab. As a result of Twitter UI changes at the end of 2022, it is now substantially more complicated to see this count for new tweets and practically impossible for tweets from 2016.
    }
    \label{fig:pol-figure}
\end{figure}

To view the private vote count before December 2022, when Twitter change its user interface (UI), a user had navigate to ``\textit{View Tweet Activity}'' from the tweet UI, then to "\textit{View all engagements}" on the subsequent menu (Figure~\ref{fig:pol-figure}). 

\section{Appendix D}
Complete significance tests between political and random group distriubtions (Table 
\ref{tab:sig-test}).

\begin{table}[h!]
\centering
\begin{tabular}{c c c} 
 \hline
 Type & Distirubtions & $p$ (MWU test) \\ 
 \hline
 Bot scores & Authors vs. Rand. Authors & $ p < 10^{-10} $ \\
 Bot scores & Retweeters vs. Rand. Rets. & $ 0.0013$ \\
 Bot scores & favoriters vs. Rand. Favs. & $ p < 10^{-10}$ \\
 Foreign & Authors vs. Rand. Authors & $ p < 10^{-10} $\\
 Foreign & Retweeters vs. Rand. Rets. & $ p < 10^{-10} $ \\
 Foreign & favoriters vs. Rand. Favs. & $ p < 10^{-10} $ \\
 Activity & Authors vs. Rand. Authors & $ p < 10^{-10} $ \\
 Activity & Retweeters vs. Rand. Rets. & $ p < 10^{-10} $\\
 Activity & favoriters vs. Rand. Favs. & $ p < 10^{-10} $ \\
 \hline
\end{tabular}
\caption{Complete significance tests by user group}
\label{tab:sig-test}
\end{table}

\section*{Appendix E}
Full list of terms to match conspiratorial polls are provided in Table \ref{search-terms}.
 
\begin{table}[ht]
\centering
\begin{tabular}{|p{0.45\textwidth}|}
\hline
\#bidenlosestotrump, \#votersuppression, accurate, accuracy, biased, blue wave crap, fraud, illegal ballots, illegal votes, illegal win, illegal winner, illegally win, illegally won, interesting statistic, legal ballots, legal votes, legal win, legal winner, legally win, legally won, real votes, real winner, recount, rejection, rejected, rigged, suppression, unbiased poll, voter suppression \\ \hline
\end{tabular}
\caption{Full list of keyword used for identifying conspiratorial polls}
\label{search-terms}
\end{table}

\end{document}